\documentstyle{camera}
\newcommand{\AmS}{{\protect\the\textfont2
  A\kern-.1667em\lower.5ex\hbox{M}\kern-.125emS}}

\newcommand{\beq}{\begin{equation}}
\newcommand{\eeq}{\end{equation}}
\newcommand{\bea}{\begin{eqnarray}}
\newcommand{\eea}{\end{eqnarray}}

\newcommand\numu{{\nu_\mu}}

\newcommand\nue{{\nu_e}}
\newcommand\anue{\bar\nu_e}

\def\dm2{\Delta m^2}
\def\sq2{sin^2(2\Theta)}

\def\nue{\nu_e}
\def\numu{\nu_\mu}

\def\anue{\bar{\nu}_e}
\def\nux{\nu_x}
\newcommand{\be}{\begin{equation}}
\newcommand{\ee}{\end{equation}}

\input epsf

\begin{document}

%%%%%%%%%%%%%%%%%%%%%%%%%%%%%%%%%%%%%%%%%%%%%%%%%%%%%%%%
% The title, all uppercase; if you want to split it in
% two or more lines, put a \\ macro at each line break
% example:
%   \title{TITLE: FIRST LINE\\ SECOND LINE}
%
\title{LVD HIGHLIGHTS}

%%%%%%%%%%%%%%%%%%%%%%%%%%%%%%%%%%%%%%%%%%%%%%%%%%%%%%%%
% The Author(S), Separated By Commas; Do not put a
% comma before the last author, use instead the \And
% macro which produces a normal ``and'' in the
% caps/small caps context
%
\author{MARCO SELVI, \\ on behalf of the LVD collaboration}

%%%%%%%%%%%%%%%%%%%%%%%%%%%%%%%%%%%%%%%%%%%%%%%%%%%%%%%%
%
\organization{Istituto Nazionale di Fisica Nucleare, Sezione di Bologna \\ 
via Irnerio, 46 - 40126 Bologna (BO) - Italy}

\maketitle

\begin{abstract}
The Large Volume Detector (LVD) in the INFN Gran Sasso National Laboratory,
Italy, is a $\nu$ observatory mainly designed to study low energy neutrinos from
the gravitational collapse of galactic objects.
The experiment has been monitoring the Galaxy since June 1992, under increasing
larger configurations: in January 2001 it has reached its final active mass $M=1$
kt. LVD is one of the largest liquid 
scintillator apparatus for the detection of stellar collapses and, 
besides SNO, SuperKamiokande and Amanda, it is a charter member of
the SNEWS network, that has become fully operational since July 1st, 2005. 
No gravitational core-collapse has been detected by LVD during 14 years of data acquisition; this allows to put an upper limit of $0.18$ events y$^{-1}$ in our galaxy at the 90$\%$ C.L.
%LVD is also equipped with a large acceptance tracking system. It 
The LVD tracking system allows the detection and the recontruction of the cosmic muon tracks in a large fraction of the whole solid angle, in particular also horizontal tracks can be reconstructed. The results of the muon depth--intensity relation and of the flux of neutrino--induced muons are presented.
Moreover, during 2006, the CNGS beam will start its operation: the performances of LVD as a beam monitor are described.
\end{abstract}
\vspace{1.0cm}

\section {The LVD experiment}

%%\subsection{The detector}

The Large Volume Detector (LVD), located in the hall A 
of the INFN Gran Sasso National
Laboratory, Italy, is a
multipurpose detector consisting of a large volume of liquid scintillator interleaved
with limited streamer tubes in a compact geometry. It has been in operation since 1992, under different increasing configurations. 
During 2001 the final upgrade took place: LVD became fully operational, 
with an active scintillator mass $M=1000$ t.

LVD consists of an array of 840 scintillator counters, 1.5 m$^3$ each,
arranged in a compact and modular geometry; each of them is viewed on the
top by three photomultipliers.  
%Up to 2004, before a re-calibration of the full detector, the counters were divided in two subsets: the external ones ($43 \%$), operated at energy threshold ${\cal E}_h\simeq 7$ MeV,and inner ones ($57 \%$), better shielded from rock radioactivity and operated at ${\cal E}_h\simeq 4$ MeV. After the re-calibration (which, started in 2004, ended during 2005) 
All the scintillation counters are operated
at a common threshold, ${\cal E}_h\simeq 5$ MeV. 
To tag the delayed $\gamma$ pulse due to
$n$-capture, all counters are equipped with an additional discrimination channel,
set at a lower threshold, ${\cal E}_l\simeq 1$ MeV.\\
Other relevant features of the detector are:
$(i)$ good event localization and muon tagging;
$(ii)$ accurate absolute and relative timing:
$\Delta\mathrm{t}_{\rm abs}=1\, \mu \rm{s}$,
$\Delta\mathrm{t}_{\rm rel}=12.5\, \rm{ns}$;
$(iii)$ energy resolution:
$\sigma_{E}/{E} = 0.07 +0.23\cdot ({E} /\rm{MeV})^{-0.5}$;
$(iv)$ very high duty cycle, i.e. $>99.5$\% in the last five years;
$(v)$ fast event recognition.

\section{Supernova neutrino physics} 
The major purpose of the LVD
experiment is the search for neutrinos from Gravitational Stellar Collapses (GSC) in our Galaxy 
%\cite{LVD}
(Aglietta et al., 1992).

\subsection{Supernova neutrino emission}
%\label{se:sn}
Indeed, in spite of the lack of a ``standard'' model of the 
gravitational collapse of a massive star, 
the correlated neutrino
emission appears to be well established. At the end of its burning phase a massive
star ($M > 8 M_{\odot}$) explodes into a supernova, originating a neutron
star which cools emitting its binding energy 
%$E_B\sim 3\cdot10^{53}$ erg 
mostly in neutrinos and antineutrinos,
%. The largest part of this energy, 
almost equipartitioned among all 
%the neutrino and antineutrino 
species:
$E_{\bar\nu_e} \sim E_{\nu_e} \sim E_{\nu_x} \sim E_B/6$ (where $\nu_x$ denotes generically
$\nu_\mu,\bar{\nu}_\mu,\nu_\tau,\bar{\nu}_\tau$ flavors). The energy spectra are
approximatively a Fermi-Dirac distribution, with different mean temperatures,
since $\nu_e,$ $\bar{\nu}_e$ and $\nu_x$ have different couplings 
with the stellar matter: $T_{\nu_e}<T_{\bar\nu_e}<T_{\nu_x}$.
In the calculations presented in this work we assume a galactic supernova explosion at a typical
distance of $D = 10$~kpc, with a total binding energy of $E_b = 3 \cdot
10^{53}$ erg and perfect energy equipartition $f_{\nu_e}=f_{\bar\nu_e}=f_{\nu_x}=1/6$.
We also assume that the fluxes of
$\nu_\mu$, $\nu_\tau$, $\bar\nu_\mu$, and $\bar\nu_\tau$ are identical; we
fix 
$T_{\nu_x} / T_{\bar{\nu}_e} = 1.5$, 
$T_{\nu_e} / T_{\bar{\nu}_e}  = 0.8$
and $T_{\bar{\nu}_e} 
%\in \{4,~7\}~ 
=5~{\rm MeV}$.

%A summary of the values of the astrophysical parameters used in our calculations is presented in table \ref{ta:param}, together with an estimation of their range of variability.
%%%, as attempted e.g. in \cite{VissProbes}. 

%\begin{table}[h!]
%\vspace{-0.3cm}
% \caption{Astrophysical parameters values used in the calculations and their assumed uncertainties. \label{ta:param}}
%\vspace{-0.3cm}
%\begin{center}
%\begin{tabular}{|l|c|c|c|}
%\hline
%Astrophysical parameter & Unit & Chosen value & Range of variability  \\
%\hline
%\hline
%$D$: distance to the star & kpc          & $10.$      & $0.2 \div 20$   \\
%\hline
%$E_b$: total energy emitted in $\nu$ & $10^{53}$ erg & $3.$  & $2. \div 5.$ \\
%\hline
%$f_{\nue}$: fraction of $E_b$ taken by $\nue$ & & $1/6$ & $1/10 \div 1/4$ \\
%\hline
%$T_{\anue}$: $\anue$--sphere temperature  & MeV & $5.$ & $4. \div 7.$ \\
%\hline
%$T_{\nu_e} / T_{\bar{\nu}_e}$ & & $0.8$ & $ 0.5 \div 0.9 $ \\
%\hline
%$T_{\nu_x} / T_{\bar{\nu}_e}$ & & $1.5$ & $ 1.1 \div 2. $ \\
%\hline
%$\eta$: pinching parameter & & $0.$ & $0. \div 2.$ \\
%\hline 
%\end{tabular}
%\end{center}
%\vspace{-0.5 cm}
%\end{table}

LVD is able to detect $\bar{\nu}_e$ interactions with protons in the scintillator, 
which give the main signal of supernova neutrinos, with a very good
signature. 
Moreover, it can detect $\nu_e$ through the elastic scattering reactions 
with electrons, $(\nue + \bar{\nu}_e)$ through charged current interactions with the
carbon nuclei of the scintillator, and it is also sensitive to neutrinos of all flavors 
detectable through neutral currents reactions with the carbon nuclei. 

The iron support structure of the detector (about $1000$ t) can also act as a 
target for electron neutrinos and antineutrinos. The products of the 
interaction can exit iron and be detected in the liquid scintillator. 

%The number of all the possible targets present in the LVD detector is listed in table \ref{ta:targ}.

%\begin{table}[h!]
%\vspace{-0.3cm}
% \caption{Number of targets in the LVD detector.}
% \label{ta:targ}
%\vspace{-0.3cm}
%\begin{center}
%\begin{tabular}{l|c|c|c}
%\hline
%Target Type & Contained in &       Mass       & Number of targets      \\
%\hline
%Free protons         & Liquid Scintillator     & $1000~t$ & $9.34~10^{31}$   \\
%Electrons     & LS & $1000~t$     & $3.47~10^{32}$  \\
%C Nuclei     & LS &    $1000~t$  & $4.23~10^{31}$      \\
%Fe Nuclei     & Support Structure & $710~t$     & $7.63~10^{30}$   \\
%\hline
%\end{tabular}
%\end{center}
%\vspace{-0.5 cm}
%\end{table}

%The amount of the detected neutrino-iron interaction can be as high as about 
%$20\%$ of the total number of interactions.
%The described features of stellar collapses are in fact common to all existing models and lead to rather
%model independent expectations for supernova neutrinos. 
%Thus, 
The signal observable in LVD, in different reactions and due to
different kinds of neutrinos, 
besides providing astrophysical informations on the nature of the collapse, is
sensitive to intrinsic $\nu$ properties, as oscillation of massive neutrinos 
and can give an important 
contribution to define some of the neutrino oscillation properties still
missing.

\subsection{Neutrino flavor transition}
\label{se:osc}
In the study of supernova neutrinos,
$\nu_{\mu}$ and $\nu_{\tau}$ are indistinguishable, both in the star and in
the detector, because of the corresponding charged lepton production threshold; consequently, in the frame of three--flavor oscillations,
the relevant parameters are just
$(\Delta m^2_{{\rm sol}}, U_{e2}^2)$ and 
$(\Delta m^{2}_{\rm atm}, U_{e3}^2)$. 
%\footnote{$U_{e1}^2=\cos^2 \theta_{13} \cdot \cos^2 \theta_{12} \simeq \cos^2 \theta_{12}$,  $U_{e2}^2=\cos^2 \theta_{13} \cdot \sin^2 \theta_{12} \simeq \sin^2 \theta_{12}$ and $U_{e3}^2 = \sin ^2 \theta_{13}$.}. 

We will adopt the following numerical values:
$\Delta m^2_{{\rm sol}}=8 \cdot 10^{-5}~{\rm eV}^2$, 
$\Delta m^{2}_{\rm atm}=2.5 \cdot 10^{-3}~ {\rm eV}^2$, 
$U_{e2}^2=0.33$ 
%\cite{VissStrum05}
(Strumia $\&$ Vissani, 2005).
%the selected solar parameters 
%$(\Delta m^2_{{\rm sol}}, U_{e2}^2)$ describe the
%LMA solution, as it results from a global analysis including solar, CHOOZ and KamLAND $\nu$ data \cite{VissStrum05}.

%As described in figure \ref{fi:crossNH} neutrinos, i
In the normal mass hierarchy (NH) scheme, neutrinos cross two so--called 
%Mikheyev--Smirnov--Wolfenstein \cite{msw} resonance layers
MSW resonance layers
in their path from the high density region where they are generated to the lower density one where they escape the star:
one at higher density (H), which corresponds to $(\Delta m^{2}_{\rm atm}, U_{e3}^2)$ 
%and $\rho=300 \div 6000~\mbox{g/cm}^3$ \footnote{the values are respectively for $E_\nu$ equal to $100$ and $5$ MeV}, 
and
the other at lower density (L), corresponding to 
$(\Delta m^{2}_{{\rm sol}}, U_{ e2}^2)$ 
%and $\rho=5 \div 100~\mbox{g/cm}^3$. 
Antineutrinos do not cross any MSW resonance.
% \cite{Panta,Dutta,Dighe}.

For inverted mass hierarchy (IH), transitions at
the higher density layer occur in the $\bar \nu$
sector, while at the lower density layer they occur 
in the $\nu$ sector. 

%Neutrinos are originated in regions of the star where the density is very high, so that the effective mixing matrix in matter is practically diagonal. Thus the created neutrino flavor eigenstate is completely projected into one neutrino mass eigenstate (represented by the thick purple line in figure \ref{fi:crossNH}). Then the neutrino starts its path through the matter to escape the star. If the matter density changes in a smooth way, then the propagation is said to be ``adiabatic''. It means that the neutrino propagates through the star being the same mass eigenstate.
%(i.e., referring to figure \ref{fi:crossNH}, staying over the same thick purple line). 
The adiabaticity condition depends both on the density variation and on the value of the oscillation parameters involved.
Given the energy range of supernova $\nu$ (up to $\sim 100~{\rm MeV}$)
and considering a star density profile $\rho \propto 1/r^3$, 
the L transition is adiabatic for any LMA solution values. Thus the probability to jump onto an adjacent mass eigenstate (hereafter called {\it flip} probability) is null  ($P_L=0$). The adiabaticity at the $H$ resonance
depends on the value of $U_{e3}^2$ in the following way 
%\cite{Dighe}
(Dighe $\&$ Smirnov, 2000)
:
$$P_{\rm H} \propto \exp~[-~ const~ U_{e3}^2~ (\Delta m^{2}_{\rm atm} / E)^{2/3}~ ]$$ where $P_{\rm H}$ is the flip probability at the H resonance.
 
When $U_{e3}^2 \geq 5 \cdot 10^{-4}$ 
the conversion is completely adiabatic ({\it ad}) and the flip probability is null ($P_{\rm H}=0$); conversely, when $U_{e3}^2 \le 5 \cdot 10^{-6}$ the conversion is completely non adiabatic ({\it na}) and the flip probability is $P_{\rm H}=1$. We used in the calculation 
$U_{e3}^2=10^{-2}$, which is just behind the corner of the CHOOZ upper limit, for the adiabatic case and $U_{e3}^2=10^{-6}$ for the non adiabatic one.

For neutrinos, in the NH-{\it ad} case $\nue$ generated in the star arrive at Earth as $\nu_3$, so their probability to be detected as $\nue$ is $U_{e3}^2 \sim 0$. Thus, the detected $\nue$ come from higher--energy $\nux$ in the star that get the Earth as $\nu_2$ and $\nu_1$. \\If the H transition is not adiabatic or if the hierarchy is inverted the original $\nue$ get the Earth as $\nu_2$ and their probability to be detected as $\nue$ is $U_{e2}^2 \sim 0.3$.

For antineutrinos, in the NH case or in the IH-{\it na}, the $\bar \nu_e$ produced in the supernova core arrive at Earth as $\nu_1$, and they have a high ($U_{e1}^2 \simeq 0.7$) probability to be detected as $\bar\nu_e$. 
On the other hand, the original $\bar \nu_x$ arrive at Earth as $\nu_2$ and $\nu_3$ and are detected as $\bar \nu_e$ with probability $U_{e2}^2$.\\ 
In the IH-{\it ad} case the detected $\bar \nu_e$ completely come from the original, higher--energy $\bar \nu_x$ flux in the star.

\begin{figure}[h!]  %%% FIGURE 1 %%%
  \begin{minipage}{.47\columnwidth}
    \epsfysize=6.5cm \hspace{4.0cm} \epsfbox{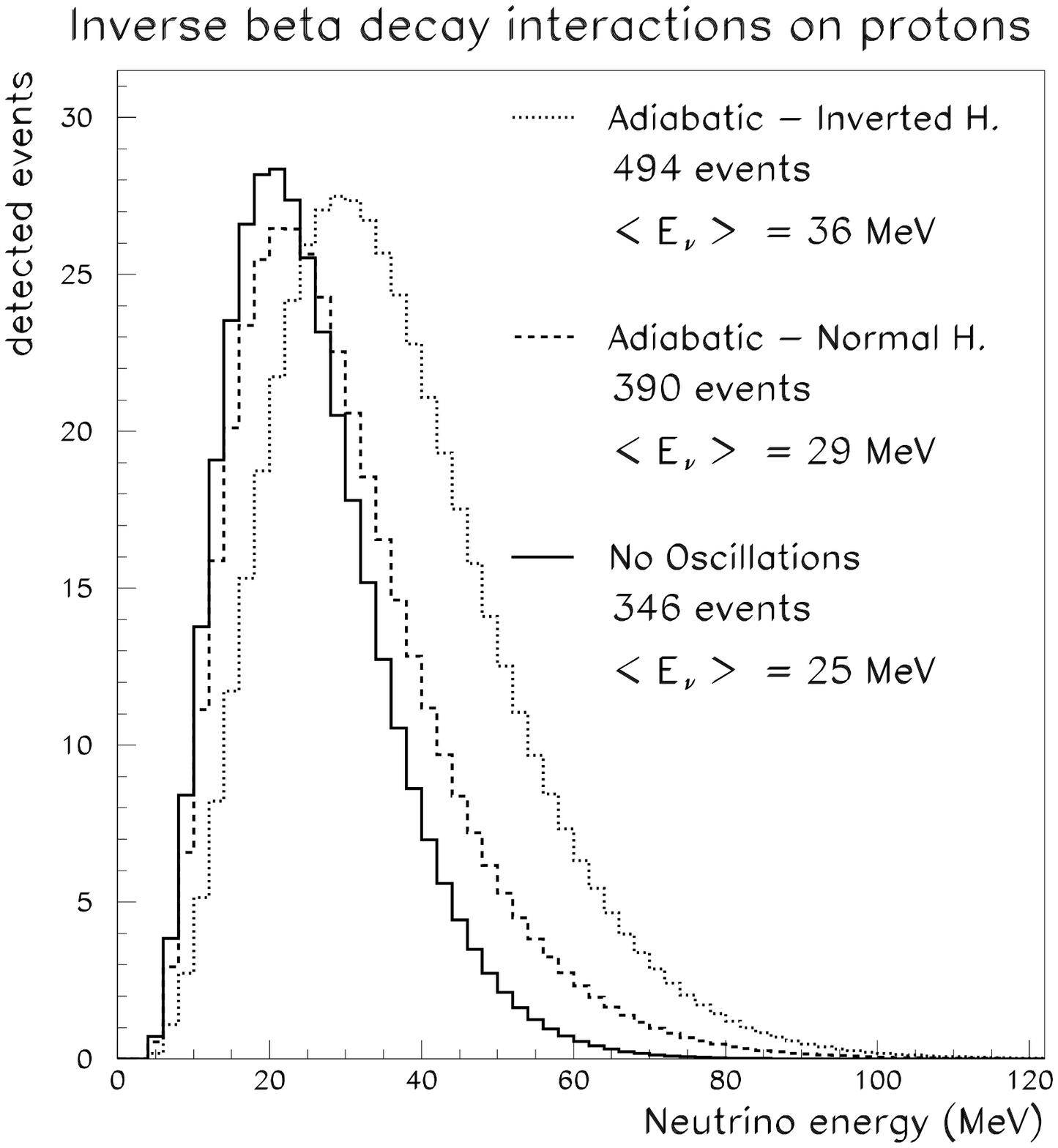}
    \vspace{-0.3cm} \caption[h]{Neutrino energy distribution in the $\bar\nu_e$ interactions with $p$ expected in LVD for three oscillation scenarios: no oscillation (solid line), adiabatic transition with NH (dashed), adiabatic transition with IH (dotted).}
    \label{fi:figibdosc}
  \end{minipage}
  \hspace{1pc} %%%%% space between two figures
  \begin{minipage}{.47\columnwidth}
    \epsfysize=6.5cm \hspace{4.0cm} \epsfbox{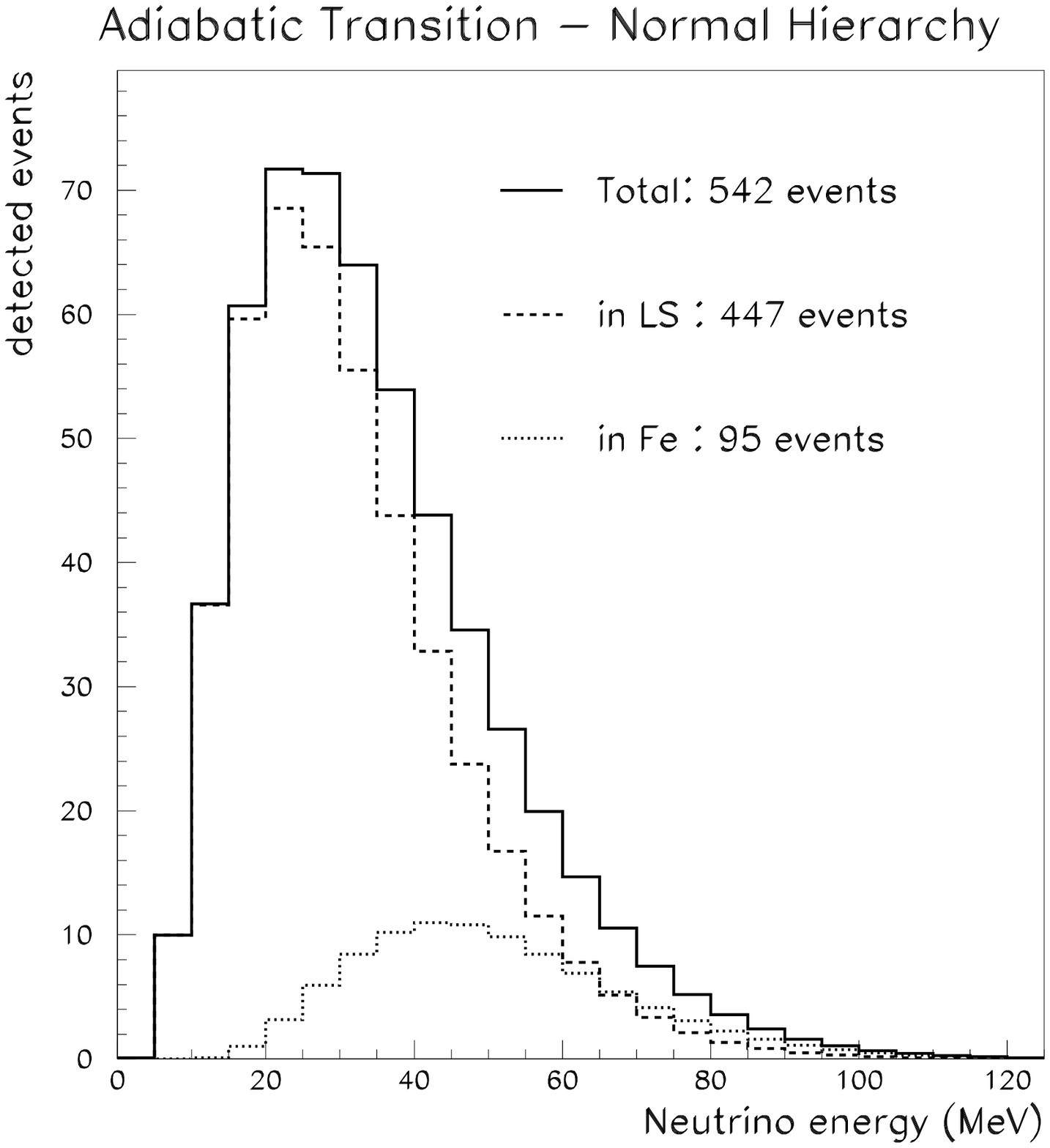}
    \vspace{-0.3cm} \caption[h]{Neutrino energy distribution of the events occurring in the liquid scintillator (dashed), in the iron support structure (dotted) and their sum (solid) in the LVD detector.}
    \label{fi:iron}
  \end{minipage}
\end{figure}

%The main interaction in LVD is the inverse beta decay (IBD) of electron antineutrinos. 
In figure \ref{fi:figibdosc} we consider the inverse beta decay of $\anue$ (the main interaction in LVD) and we show the energy spectra of the detected $\anue$ in the case of no oscillation and in the case of adiabatic transition with NH and IH. 
%We remind here that the non--adiabatic transition case (for both NH and IH) is coincident with the adiabatic NH case.
In the case of oscillation, adiabatic, normal hierarchy, there is a contribution ($\sin^2\theta_{12}$) of the originally higher--energy $\bar \nu_x$ which gives rise to a higher average neutrino energy and to a larger number of detected events.
The $\nu_x$ contribution is even higher ($\sim 1$) if the transition is adiabatic and the hierarchy inverted, because the MSW resonance happens in the $\bar\nu$ sector. 
%This results in a  higher neutrino energy, as visible in figure  \ref{fi:figibdosc}, and in a larger number of events.

Another important contribution to the total number of events is also given by neutrino interactions in the iron support structure of the LVD detector. Given the rather high effective threshold (about $10$ MeV) and the increasing detection efficiency with the neutrino energy, they are concentrated in the high energy part of the spectrum ($E_\nu > 20$ MeV). 
%Thus they are extremely sensitive to the neutrino energy spectrum and, indeed, to the oscillation parameters.
%In figure \ref{fi:repfe} we show the dependence of the total number of detected $(\nu_e + \bar \nu_e)$ CC interactions with Fe to the $\anue$--sphere temperature 
%, in the various oscillation scenarios.
In figure \ref{fi:iron} we show the contribution of $(\nu_e+\bar \nu_e)$ {\rm Fe} interactions on the total number of events. For the chosen supernova and oscillation parameters they are about $17\%$ of the total signal.

The expected number of events in the various LVD detection channels and in the different oscillation scenarios
%and the mean energy of the detected $\anue~p$ events 
are shown in table \ref{ta:central}.

\begin{table}[h!] %checked
\vspace{-0.3 cm}
 \caption{Expected results in the various LVD detection channels and in the mean energy of the detected $\anue~p$ events.
%, calculated considering the chosen values of the astrophysical parameters, as given in table \ref{ta:param}.
\label{ta:central}}
\vspace{-0.3 cm}
\begin{center}
\begin{tabular}{|l|c|c|c|c|}
\hline
  & No Oscillation & Non Adiabatic & Adiabatic NH & Adiabatic IH  \\
\hline
\hline
$\anue ~ p$   & 346. & \multicolumn{2}{c|}{391.} &   494. \\
\hline
$\langle E_{\anue} \rangle$ in $\anue ~ p$ & 25. MeV  & \multicolumn{2}{c|}{30. MeV}  & 37. MeV \\
\hline
 CC with $^{12}$C & 8. & 22. &   29. &    27. \\
\hline
 CC with $^{56}$Fe & 22. & 72. &   95. &    92. \\ 
\hline
 NC with $^{12}$C  & \multicolumn{4}{c|}{27}  \\
\hline
\end{tabular}
\end{center}
\vspace{-0.5 cm}
\end{table}

%%%%%%%%%%%%%%%%%%%%%%%%%%%%%%%%%%%%%%%%%%%%%%%%%%%%%%%%%%%%%%%%
%%%%%%%%%%%%%%%%%%%%%%%%%%%%%%%%%%%%%%%%%%%%%%%%%%%%%%%%%%%%%%%%

%%%%\subsection {Supernova physics}

\subsection{Monitoring the Galaxy}
\begin{figure}[h!]  %%% FIGURE 3 e 4 %%%
  \begin{minipage}{.47\columnwidth}
    \epsfysize=5.5cm \hspace{4.0cm} \epsfbox{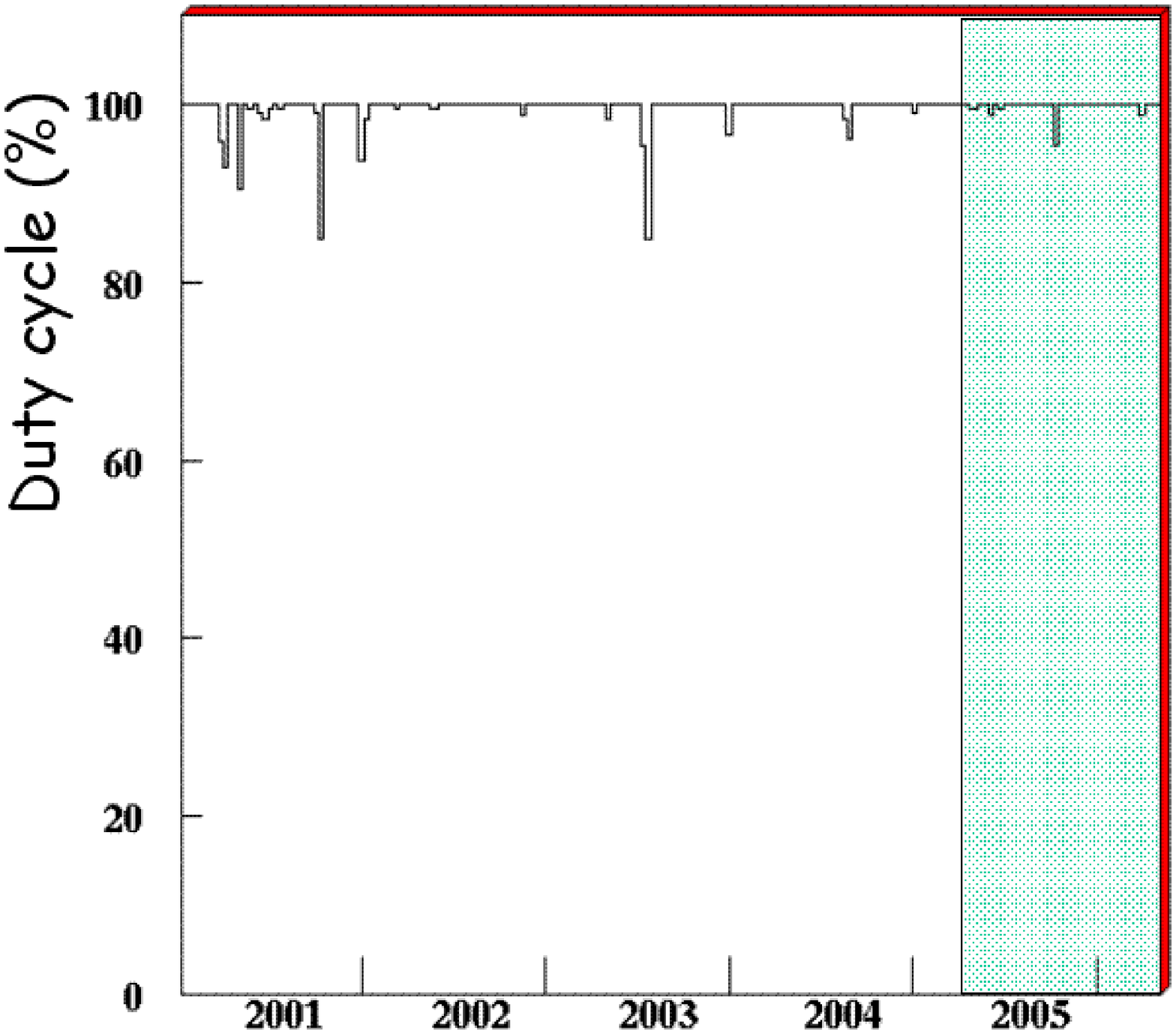}
    \vspace{-0.3cm} \caption{LVD duty cycle during the last 5 years of data acquisition.}
    \label{fi:duty}
  \end{minipage}
  \hspace{1pc} %%%%% space between two figures
  \begin{minipage}{.47\columnwidth}
    \epsfysize=5.5cm \hspace{4.0cm} \epsfbox{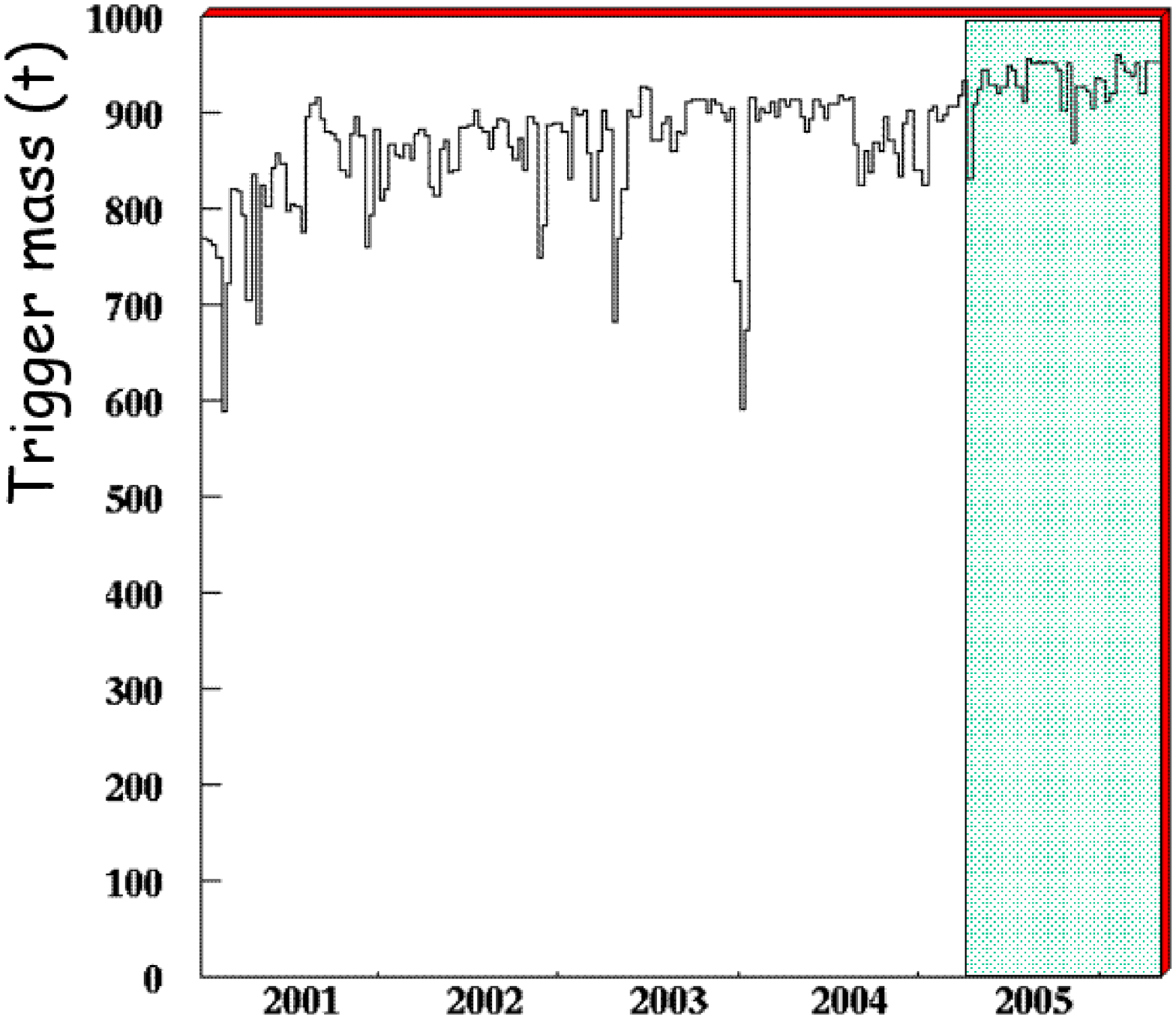}
    \vspace{-0.3cm} \caption{LVD active mass during the last 5 years of data acquisition.}
    \label{fi:actmass}
  \end{minipage}
\end{figure}

LVD has been continuously monitoring the Galaxy since 1992 in the search for
neutrino bursts from GSC
\footnote{The results of this search
have been periodically updated and
published in the ICRC Proceedings, since 1993 until 2005.
%\cite{ICRC93,ICRC95,ICRC97,ICRC99,ICRC01,ICRC03,ICRC05} 
}.
Its active mass has been progressively increased from about $330$
t in 1992 to $1000$ t in 2001, always guaranteeing a
sensitivity to gravitational stellar collapses 
up to distances $d=20$ kpc from the Earth, even
in the case of the lowest $\nu$-sphere temperature.\\[0.2cm]
%The telescope duty cycle has been continuously improving since 1992.
In fig. \ref{fi:duty} and \ref{fi:actmass} we show respectively the duty cycle and the average active mass, during the last 5 years. Considering just the last year (shaded areas) the average duty cycle
was $99.98\%$ and the average active mass $940$ t. 

%%%%%%%%%%%%%%%%%%%%%%%%%%%%%%%%%%%%%%%%%%%%%%%%%%%%%%%%%%%%%%%%
%%%%%%%%%%%%%%%%%%%%%%%%%%%%%%%%%%%%%%%%%%%%%%%%%%%%%%%%%%%%%%%%

%The counting rate is continuously monitored and a
All events are processed on the base of their time sequence, searching for cluster not compatible with a poissonian fluctuation of the background. A first selection consists in rejecting the crossing muon events and accepting only those with energy in the $[7 - 100]$ MeV range. After this selection cut we obtain a very stable counting rate $f_b=0.3$ Hz.
Each cluster is identified by its multiplicity $m$ and its duration $\Delta t$ (up to $200$ s): the imitation frequency $Fim$ (that is the frequency that a particular cluster ($m$,$\Delta t$) is due to a poissonian fluctuation of the background $f_b$) is then calculated. The interesting candidates (those with $Fim < 1$ ev/year) undergo several consistency checks: (i) the topological distribution of pulses inside the detector has to be uniform; (ii) their energy spectrum is checked against the background spectrum 
%compared with the thermal Fermi-Dirac expectation 
and (iii) the time distribution of the delayed low energy pulses (due to the $2.2$ MeV gamma from the neutron capture) must follow an exponential law.

\noindent No significant signal has been registered by LVD during 14 years of data acquisition.
Since the LVD sensitivity is higher than what is expected from GSC models (even if
the source is at a distance of 20 kpc and for soft neutrino energy spectra)
we can conclude that no gravitational stellar collapse has occurred in the
Galaxy in the whole period of observation: the resulting upper limit to the
rate of GSC, updated to April, 2006, at 90\% C.L. is 0.18 events/yr.

\subsection {SNEWS}

The SNEWS (SuperNova Early Warning System)
%\cite{Snews} 
(Antonioli et al., 2004)
project 
is an international collaboration including several experiments 
sensitive to a core-collpase supernova neutrino signal in the Galaxy and neighbour.
The goal of SNEWS is to provide the astronomical community with a prompt 
and confident alert of the occurrence of a Galactic supernova event, generated
by the coincidence of two or more active detectors.
In addition the collaboration is engaged in cooperative work, such as 
the downtime coordination and inter-experiment timing verification, 
designed to optimize the global sensitivity to
a supernova signal.\\
A dedicated process waits for alarm datagrams from the experiments' clients, and provides an alert if there is a coincidence within a specified time window
(10 seconds for normal running). 
In July 2005, after a few years of tuning, the charter members of 
SNEWS (i.e., LVD, Super-K and SNO) together with the newly joined
Amanda/IceCube, started the effective operation of the network, which means that the alert is really sent to the list subscribers, in the case of an at least two-fold coincidence (see {\it snews.bnl.gov} to get your own SN alert !). \\
%There is currently a single coincidence server, hosted by Brookhaven National Laboratory. 
%The BNL computer continuously runs a coincidence server
%A scheme of ``GOLD'' and ``SILVER'' alerts
%has been implemented: GOLD alerts are automatically disseminated
%to the community; SILVER alerts are disseminated only after human checking.\\
Up to now, no inter-experiment coincidence, real or accidental, has ever
occurred (except during a special high rate test mode), nor any core collapse
event been detected within the lifetimes of the currently active
experiments.

\section{Cosmic Ray Physics}
%LVD is composed of 3 identical towers. 
The scintillation counters are interleaved by a large acceptance tracking system made of limited streamer tubes ($1 \times 1 cm^2$ cell' cross section): two staggered layers cover the bottom and one lateral side of the cluster of 8 scintillation counters. They are read out by $4~cm$ strips.
%Each LVD tower contains 38 identical modules consisting of eight scintillation counters.
%The modules are interleaved by a large acceptance tracking system made of limited streamer tubes, which cover the bottom and one side of the cluster of 8 counters. 
The tracking system allows the detection and the recontruction of the cosmic muon tracks in a large fraction of the whole solid angle, in particular also horizontal tracks can be reconstructed. The total acceptance of one LVD tower is about $350~m^2~sr$.
The tracking system has been operative since the beginning of the experiment (1992) until 2002. Seven million muon events have been detected and reconstructed during about $70000~h$ of live time for data acquisition.
In the analysis we have used muon events with all multiplicities. The acceptances for each angular bin have been calculated using the simulation of muons passing through LVD taking into account muon interaction with the detector material and the detector response. As a result of the data processing the angular distribution of the number of detected muons $N_{\mu}(\phi,\cos \theta)$ has been obtained. The angular informations allow to derive the amount of rock crossed by the muons, given the knowledge of the Gran Sasso mountain profile.
Thus, the so--called depth--intensity relation has been calculated, for a very large range of slant depths ($[3 - 20]$ km $w.e.$), see figure \ref{fi:muons}. Two main components can be identified. The first one is dominant at slant depth up to $13$ km $w.e.$ and is due to conventional cosmic muons, i.e. high energy downward muons produced by $\pi$ and $K$ mesons in the atmosphere.
When fitted with the formula 
$I_{\mu}(x) = A \left ( \frac{x_0}{x} \right )^\alpha \exp^{-x/x_0}$ we obtain: $A= (2.17^{+0.12}_{-0.18}) ~ 10^{-6}~cm^{-2} sr^{-1} s^{-1}$, $x_0 = (1175 \pm 10)~ hg/cm^2$ and $\alpha = 2.07 \pm 0.03$. Those results are in agreement with those of the other underground experiments. Previous results were published in (Aglietta et al., 1998).
%\cite{LVDmu} 
The second is due to horizontal muons produced by neutrino interaction in the rock surrounding LVD. They are independent of the slant depth and thus they become observable when the atmospheric muon flux is suppressed by the large amount of rock. Their value is $(4.95 \pm 1.15_{stat} \pm 0.4_{syst} ) ~ 10^{-13}~cm^{-2} sr^{-1} s^{-1}$.

\section{CNGS beam monitor}
\begin{figure}[h!]  %%% FIGURE 5 e 6 %%%
  \begin{minipage}{.47\columnwidth}
    \epsfysize=6.5cm \hspace{4.0cm} \epsfbox{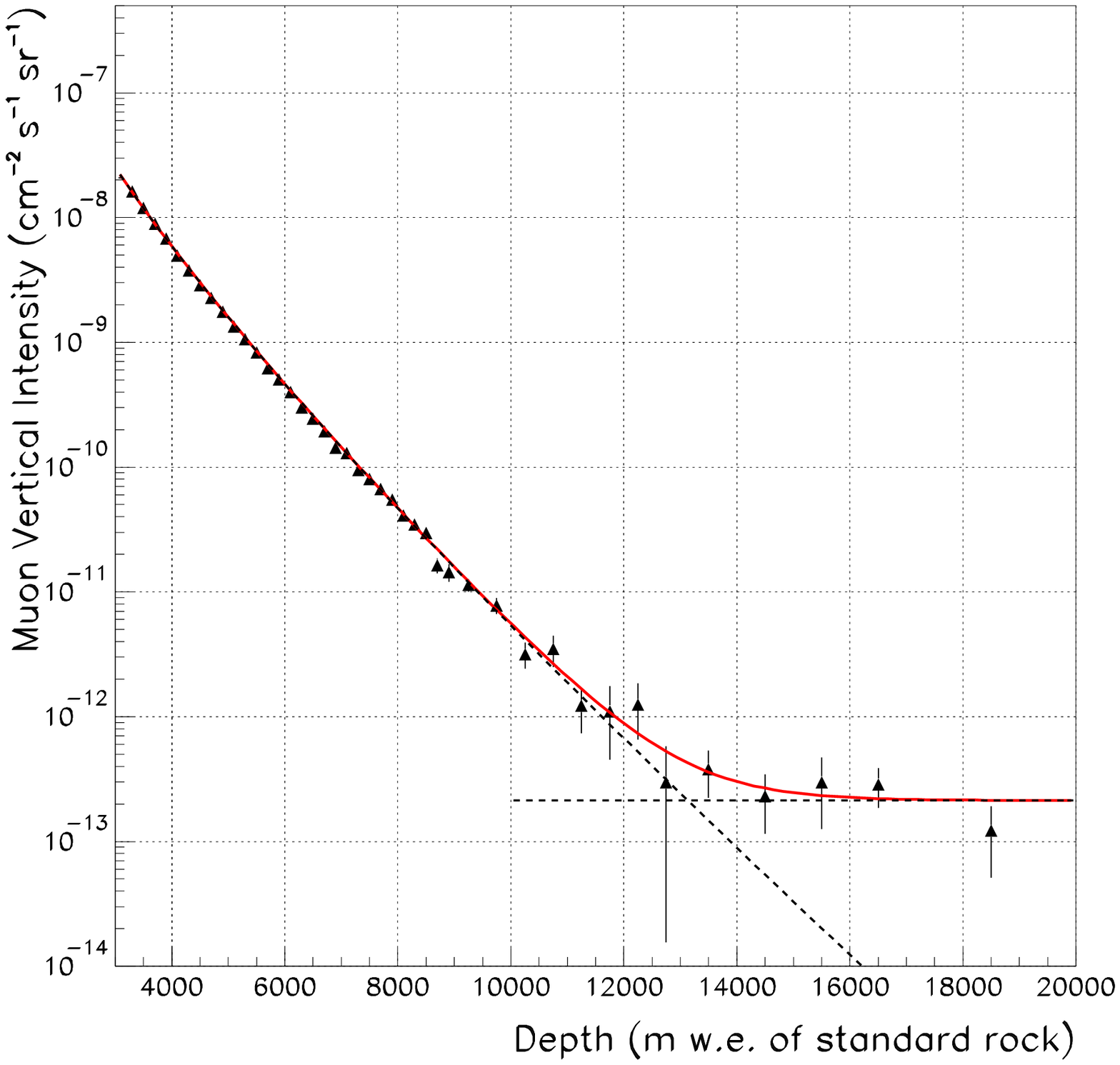}
    \vspace{-0.3cm} \caption{Vertical muon intensity as a function of the traversed slant depth in standard rock.} 
    \label{fi:muons}
  \end{minipage}
  \hspace{1pc} %%%%% space between two figures
  \begin{minipage}{.47\columnwidth}
    \epsfysize=6.5cm \hspace{4.0cm} \epsfbox{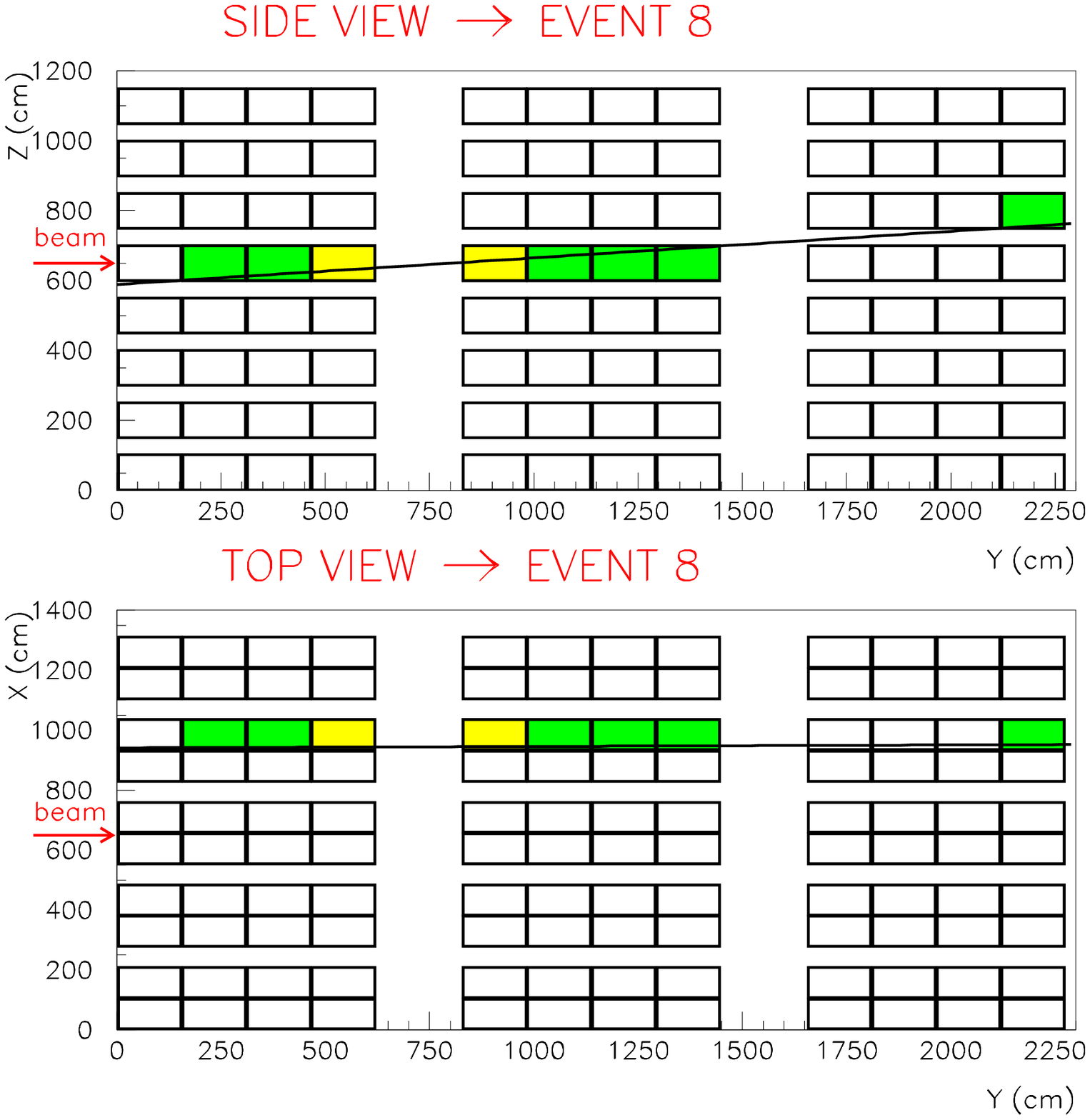}
    \vspace{-0.3cm} 
\caption{Top and side view of a typical CNGS event inside LVD: a muons originated by a CC $\numu$ interaction in the rock upstream the LNGS.}
    \label{fi:cngs}
  \end{minipage}
\end{figure}

The CNGS beam will start its operation during the summer 2006. It is a high energy $\numu$ beam mainly devoted to study neutrino oscillation through the appearance of $\nu_{\tau}$ at the LNGS. The LVD detector will act as a useful beam monitor 
(Aglietta et al., 2004).
%\cite{LVDcngs}
Two main kind of events will be detected: muons originated through charged current neutrino interaction in the rock upstream of the LNGS (about $120$ per days at nominal beam intensity, see a typical events in figure \ref{fi:cngs}) and the charged and neutral current interaction inside the apparatus (about $30$ per day). The background due to cosmic muons is negligible because of the clear time structure of the CNGS beam spill and the orientation of the events (mainly vertical for cosmics, horizontal for the CNGS ones).
In one week of data acquisition a $3 \%$ statistical accuracy can be obtained, useful to check for the overall beam orientation. 
%Moreover, during the first weeks of beam operation, LVD will be the only active detector at Gran Sasso able to detect the CNGS events, thus its role will be particularly important.

%%%%%%%%%%%%%%%%%%%%%%%%%%%%%%%%%%%%%%%%%%%%%%%%%%%%%%%%%%%%%%%%
%%%%%%%%%%%%%%%%%%%%%%%%%%%%%%%%%%%%%%%%%%%%%%%%%%%%%%%%%%%%%%%%

%\begin{figure}[t!]
%      \vspace{10truecm}
%\special{psfile=M51_LARGE_SCALE_B.ps  voffset=140 hoffset=0
%hscale=30.0 vscale=18.7 angle=0}
%\special{psfile=M83_LARGE_SCALE_B.ps voffset=141 hoffset=220
%hscale=18.0 vscale=18.5 angle=0}
%\special{psfile=NGC4631_B_2-8cm.ps voffset=-8 hoffset=35
%hscale=18.3 vscale=20 angle=0} \special{psfile=M82_B_VLA.ps
%voffset=0 hoffset=220 hscale=18 vscale=18.2 angle=0}
%      \caption[h]{The magnetic field structures of different
%galaxies. Clockwise from the left upper panel: orientation of the
%large-scale magnetic fields in M51, orientation of the large-scale
%magnetic fields in M83 (Neininger et al., 1991), magnetic fields
%in NGC 4631 from 2.8 cm observations (Wielebinski \& Krause,
%1993), and magnetic fields in M82 from VLA observations (Reuter et
%al., 1993).

%{\bf The use of the commands for this composed figures is rather
%simple. It is necessary to vary "voffset" and "hoffset" in order
%to move each figure in vertical and in horizontal, respectively.
%It is necessary to vary "hscale" and "vscale" in order to resize
%each figure in horizontal and in vertical, respectively. With the
%command "angle" it is possible to rotate a figure.}}
%     \label{fig1}
%    \end{figure}

%\section{Acknowledgements}
%This work was partially supported by the European Community. Many thanks
%to Drs Black and White for useful discussions.

%\bigskip
\bigskip
\noindent {\bf DISCUSSION}

%\bigskip
\vspace{0.2cm}
\noindent {\bf JIM BEALL:} What distance did you use in the estimate of hundreds of events for a galactic SN ?

%\bigskip
\vspace{0.1cm}
\noindent {\bf MARCO SELVI:} We considered a SN core collapse at $10$ kpc, close to the center of the galaxy ($8.5$ kpc). 
%We expect hundreds of events in the whole LVD detector, summing up all the various detection channels.

%\bigskip
\vspace{0.1cm}
\noindent {\bf JIM BEALL's Comment:} I would like to underline
that the continuous and continued operation of this detector is very important.

\end{document}